\documentclass[10pt,conference]{IEEEtran}
\IEEEoverridecommandlockouts

\IfFileExists{cite.sty}{\usepackage{cite}}{}
\usepackage{amsmath,amssymb,amsfonts}
\usepackage{graphicx}
\IfFileExists{subfig.sty}{%
  \usepackage[caption=false,font=footnotesize]{subfig}%
}{%
}
\usepackage{textcomp}
\usepackage{xcolor}

\usepackage{tikz}
\usetikzlibrary{calc,arrows.meta,shapes.geometric}

\usepackage{tikz}
\usetikzlibrary{arrows.meta, positioning, calc, shapes.geometric, shapes.arrows, fit, backgrounds, decorations.pathreplacing, matrix}

\definecolor{d4cBlue}{HTML}{2B5DAA}
\definecolor{d4cTeal}{HTML}{1B9E77}
\definecolor{d4cDark}{HTML}{1F2937}
\definecolor{d4cLight}{HTML}{F3F4F6}

\tikzset{
  >=Latex,
  d4cTitle/.style={font=\bfseries, text=d4cDark},
  d4cLabel/.style={font=\footnotesize, text=d4cDark, align=center},
  d4cSmall/.style={font=\footnotesize},
  d4cBox/.style={draw=d4cDark, thick, rounded corners=2pt, fill=white, inner sep=5pt, align=left},
  d4cGroup/.style={draw=d4cBlue, thick, rounded corners=3pt, inner sep=6pt},
  d4cGroupDashed/.style={d4cGroup, dashed},
  d4cLine/.style={thick, draw=d4cDark},
  d4cArrow/.style={-Latex, thick, draw=d4cDark},
  d4cArrowTeal/.style={-Latex, thick, draw=d4cTeal},
  d4cArrowBlue/.style={-Latex, thick, draw=d4cBlue},
  d4cArrowRedDashed/.style={-Latex, thick, draw=red!70!black, dashed},
}

\tikzset{
  d4cCabin/.style={draw=d4cDark, thick, rectangle, minimum width=10mm, minimum height=6mm, fill=white, font=\footnotesize},
  d4cInverter/.style={
    draw=d4cDark, thick, rectangle, minimum size=10mm, fill=white,
    path picture={
      \draw[d4cDark, thick] (path picture bounding box.south west) -- (path picture bounding box.north east);
      \node[font=\tiny, anchor=north west, inner sep=1pt] at (path picture bounding box.north west) {$\sim$};
      \node[font=\tiny, anchor=south east, inner sep=1pt] at (path picture bounding box.south east) {$=$};
    }
  },
  d4cLoad/.style={draw=d4cDark, thick, isosceles triangle, isosceles triangle apex angle=60, shape border rotate=180,
                  minimum width=8mm, minimum height=8mm, fill=white},
  d4cTransformer/.style={
    draw=d4cDark, thick, circle, minimum size=9mm, fill=white,
    path picture={
      \draw[d4cDark, thick] (path picture bounding box.south west) -- (path picture bounding box.north east);
      \draw[d4cDark, thick, -Latex] ($(path picture bounding box.north east)+(-2pt,-2pt)$) -- ++(4pt,4pt);
    }
  },
  d4cSwitch/.style={draw=d4cDark, thick, rectangle, rounded corners=2pt, fill=d4cLight,
                   minimum width=18mm, minimum height=8mm, font=\footnotesize, align=center},
  d4cVM/.style={draw=d4cTeal!60!black, thick, rectangle, rounded corners=2pt, fill=d4cTeal!8,
               minimum width=18mm, minimum height=8mm, font=\footnotesize, align=center},
  d4cAttacker/.style={draw=red!70!black, thick, rectangle, rounded corners=2pt, fill=red!6,
                     minimum width=18mm, minimum height=8mm, font=\footnotesize, align=center},
}

\tikzset{
  d4cStageArrow/.style={
    single arrow,
    single arrow head extend=5mm,
    shape border rotate=0,
    draw=none,
    fill=d4cBlue,
    minimum height=9mm,
    inner sep=6pt,
    text=white,
    font=\bfseries
  },
  d4cStageBox/.style={
    draw=d4cBlue, thick,
    fill=white,
    inner sep=6pt,
    align=left,
    font=\scriptsize,
  },
}

\usepackage{booktabs}
\usepackage{array}
\usepackage{tabularx}

\IfFileExists{algorithm.sty}{\usepackage{algorithm}}{}
\IfFileExists{algorithmic.sty}{\usepackage{algorithmic}}{}
\IfFileExists{dblfloatfix.sty}{\usepackage{dblfloatfix}}{}
\IfFileExists{balance.sty}{\usepackage{balance}}{}
\IfFileExists{placeins.sty}{\usepackage{placeins}}{}

\newcolumntype{L}[1]{>{\raggedright\arraybackslash}p{#1}}

\usepackage[hyphens]{url}
\usepackage[hidelinks]{hyperref}

\DeclareRobustCommand{\icode}[1]{\texttt{\detokenize{#1}}}
\newcommand{\email}[1]{\nolinkurl{#1}}
\newcommand{\fcode}[1]{\path{#1}}

\begin{document}

\title{Data4Cyber: Labeled Cyber-Physical Datasets for Distribution-Grid Management Systems Under Representative OT Attacks}

\author{%
\IEEEauthorblockN{%
\"{O}mer Sen\IEEEauthorrefmark{1},
Lennart Bader\IEEEauthorrefmark{2},
Marawan Emara\IEEEauthorrefmark{2},
Martin Serror\IEEEauthorrefmark{2},
Benedikt Schwalm\IEEEauthorrefmark{1},\\
Junichi Tsurumi\IEEEauthorrefmark{4},
Tatsumi Oba\IEEEauthorrefmark{3},
Tomohiro Izawa\IEEEauthorrefmark{3},
Yoshihiro Ujiie\IEEEauthorrefmark{3}%
}
\IEEEauthorblockA{\IEEEauthorrefmark{1}Fraunhofer FIT,
Sankt Augustin, Germany\\
Emails: \email{{oemer.sen, benedikt.schwalm}@fit.fraunhofer.de}}
\IEEEauthorblockA{\IEEEauthorrefmark{2}Fraunhofer FKIE,
Wachtberg, Germany\\
Emails: \email{{lennart.bader, marawan.emara, martin.serror}@fkie.fraunhofer.de}}
\IEEEauthorblockA{\IEEEauthorrefmark{3}Panasonic, Japan;
\IEEEauthorrefmark{4}Panasonic, Germany\\
Emails: \email{junichi.tsurumi@eu.panasonic.com},
\email{{izawa.tomohiro, ujiie.yoshihiro, oba.tatsumi}@jp.panasonic.com}}
}

\maketitle

\begin{abstract}
Cyberattacks against power systems increasingly exploit industrial communication and control dependencies arising from distributed energy resources (DER), remote management, and market-based incentives. Progress on intrusion detection is constrained by a lack of public datasets that align operational technology (OT) network telemetry with physical measurements and control context. \emph{Data4Cyber} is a labeled cyber-physical dataset recorded in a laboratory distribution management system (DMS) for self-consumption optimization with photovoltaic generation, battery storage, controllable loads, and substation metering. It captures synchronized process measurements, controller context, price incentives, and packet captures (Modbus/TCP and MQTT) across one benign baseline and six attack scenarios spanning protocol-semantic Modbus manipulation, man-in-the-middle (MitM) false-data injection, and MQTT price manipulation. Its fixed 1\,s time base, phase-resolved labels, and state-oriented representation support temporally blocked stateful-detector evaluation. Baseline results show that high recall alone is insufficient---several detectors achieve sensitivity through alarm saturation---and that context-driven attacks require process-aware interpretation rather than single-channel thresholding.
\end{abstract}


\section{Introduction}
Digitalization of distribution grids and behind-the-meter energy management expands the attack surface of power systems. Distributed energy resources (DER), such as photovoltaic (PV) inverters and battery storage systems (BSS), are frequently integrated via Internet Protocol (IP)-based operational technology (OT) networks and protocols with limited built-in authentication and integrity protection~\cite{r1}. In parallel, operation increasingly depends on data-driven optimization and coordination signals, such as dynamic prices exchanged between edge controllers and higher-level services~\cite{r3}. These developments create tightly coupled cyber-physical dependencies where manipulation of measurements, commands, or incentives can induce unsafe operation or economic harm.

Real-world incidents demonstrate that OT-capable adversaries can achieve operational impact. The 2015 Ukraine outage and later events associated with Industroyer/CrashOverride show how attackers can leverage industrial protocol semantics and operator workflows~\cite{r10,r11}. Motivated by such threats, industrial intrusion detection systems (IDSs) and process-aware anomaly detection have seen rapid growth. However, systematic evaluation remains constrained by the scarcity of datasets that align network observations with process ground truth and control context~\cite{r13}. In particular, datasets reflecting DER-centric distribution management systems (DMSs) and including both OT traffic (e.g., Modbus/TCP) and higher-level messaging (e.g., MQTT price signals) remain limited~\cite{r20}.

This work develops and evaluates a dataset structure that links cyber events to physical consequences and controller-relevant context in DER-rich DMS settings. The contributions are: (i)~a cyber-physical laboratory testbed and multi-layer instrumentation for joint network, control, and physical analysis; (ii)~a scenario suite with one benign baseline and six attack scenarios spanning Modbus and MQTT manipulation; (iii)~a synchronized multi-layer data representation with phase-resolved labels and per-step state descriptions for stateful cyber-physical IDS evaluation; and (iv)~a standardized baseline evaluation that demonstrates dataset difficulty, temporal-generalization risks, and the need for process-aware detection. Sec.~II positions the dataset against prior work, Sec.~III details the testbed and dispatch logic, Sec.~IV defines the attack catalog, Sec.~V describes the dataset and reports the baseline evaluation, and Sec.~VI concludes.

\section{Background and Related Work}

\subsection{DMS and ICT Coupling}
Distribution management increasingly integrates DER assets (e.g., PV inverters, BSS units, controllable loads) and relies on remote monitoring and control to optimize self-consumption, peak shaving, and grid support~\cite{r3}. In typical deployments, supervisory controllers obtain telemetry from meters and inverters and issue setpoints to DER devices. This information and communication technology (ICT) coupling is frequently realized via Modbus/TCP for polling and setpoint writes~\cite{r6}. High-level coordination and incentive signals often use publish/subscribe messaging such as Message Queuing Telemetry Transport (MQTT)~\cite{r8}. While ICT coupling improves flexibility and efficiency, it also increases exposure to network-based threats and integrity violations.

\subsection{Cybersecurity of DER Control and OT Networks}
OT networks commonly operate under stringent availability constraints and rely on protocols without modern authentication and integrity protections~\cite{r2}. Consequently, threat models for DER management include reconnaissance and service discovery, unauthorized Modbus access, man-in-the-middle manipulation via ARP spoofing~\cite{r9}, and manipulation of non-physical information channels such as price incentives~\cite{r8}. Industroyer/CrashOverride illustrates a broader class of OT malware and adversarial tooling using protocol knowledge to achieve physical impact~\cite{r12}.

\subsection{Related Work: ICS and Smart-Grid Datasets}
Existing industrial-control-system (ICS) datasets cover single domains or limited-scale testbeds. The water and process families dominate. SWaT~\cite{r15}, WADI~\cite{r16}, and HAI~\cite{r17} record physical and network traces from water-treatment, water-distribution, and hardware-in-the-loop process plants with curated attack campaigns. They are well suited for continuous-process control loops, but they do not expose the dispatch logic, inverter dynamics, or protocol mix of distribution-grid management, and none of them couples Modbus/TCP measurement traffic with a higher-level publish/subscribe channel for incentive signals.

Power-domain datasets target adjacent settings rather than DER-centric distribution management. PowerDuck~\cite{r18} records substation IEC~61850 Generic Object Oriented Substation Event (GOOSE) traffic but excludes inverter setpoints, battery state-of-charge dynamics, and economic incentive channels. Sherlock~\cite{r20} and adjacent co-simulation environments for active distribution grids~\cite{r23} broaden the attack catalog but do not exercise real DER hardware or vendor-specific Modbus semantics. Survey work~\cite{r13} likewise identifies the bottleneck as the absence of datasets that align network observations with process ground truth and controller context in realistic operational settings. Table~\ref{tab:related_positioning} compares Data4Cyber against representative datasets along the criteria that matter for process-aware IDS research: physical hardware, DER assets, protocol mix, controller context, phase labels, and reusable data products. Data4Cyber closes the gap by combining real DER hardware, mixed Modbus/TCP and MQTT traffic, synchronized physical metering, and an MQTT-based price-signal manipulation class that, to our knowledge, is not represented in publicly available smart-grid datasets.

\subsection{FAIR and Reusable Security Datasets}
For reproducible IDS benchmarking, data availability alone is insufficient; variables, labels, timing, and scenario semantics must also be interpretable without laboratory-specific knowledge. Data4Cyber therefore follows the FAIR principles---findable, accessible, interoperable, and reusable~\cite{r22}. Findability and accessibility are achieved through a persistent archive with stable metadata; interoperability through machine-readable CSV, JSON, PCAP, and state-log views; and reusability through explicit scenario descriptions, semantic mappings, labels, and documented evaluation settings. These choices address a recurring limitation in security datasets: traces may be downloadable yet difficult to align, interpret, or feed into existing IDS pipelines.

\begin{table}[!t]
\caption{Positioning of Data4Cyber relative to representative datasets.}
\label{tab:related_positioning}
\centering
\footnotesize
\setlength{\tabcolsep}{3pt}
\renewcommand{\arraystretch}{1.08}
\begin{tabular}{L{1.35cm} L{1.55cm} L{1.35cm} L{2.45cm}}
\toprule
\textbf{Dataset} & \textbf{Domain} & \textbf{Primary Scope} & \textbf{Remaining Gap Addressed}\\
\midrule
SWaT\slash WADI & Water cyber-physical & Process + network & DER-centric DMS with grid assets and price/control context.\\
HAI & ICS hardware-in-the-loop & Process + attack labels & Adds OT PCAPs, Modbus/MQTT semantics, and DMS-specific attacks.\\
PowerDuck & Substation & IEC~61850 traffic & Targets DER management rather than substation GOOSE only.\\
Sherlock & Power grid & Co-simulation & Uses laboratory DER assets and synchronized physical measurements.\\
\bottomrule
\end{tabular}
\end{table}

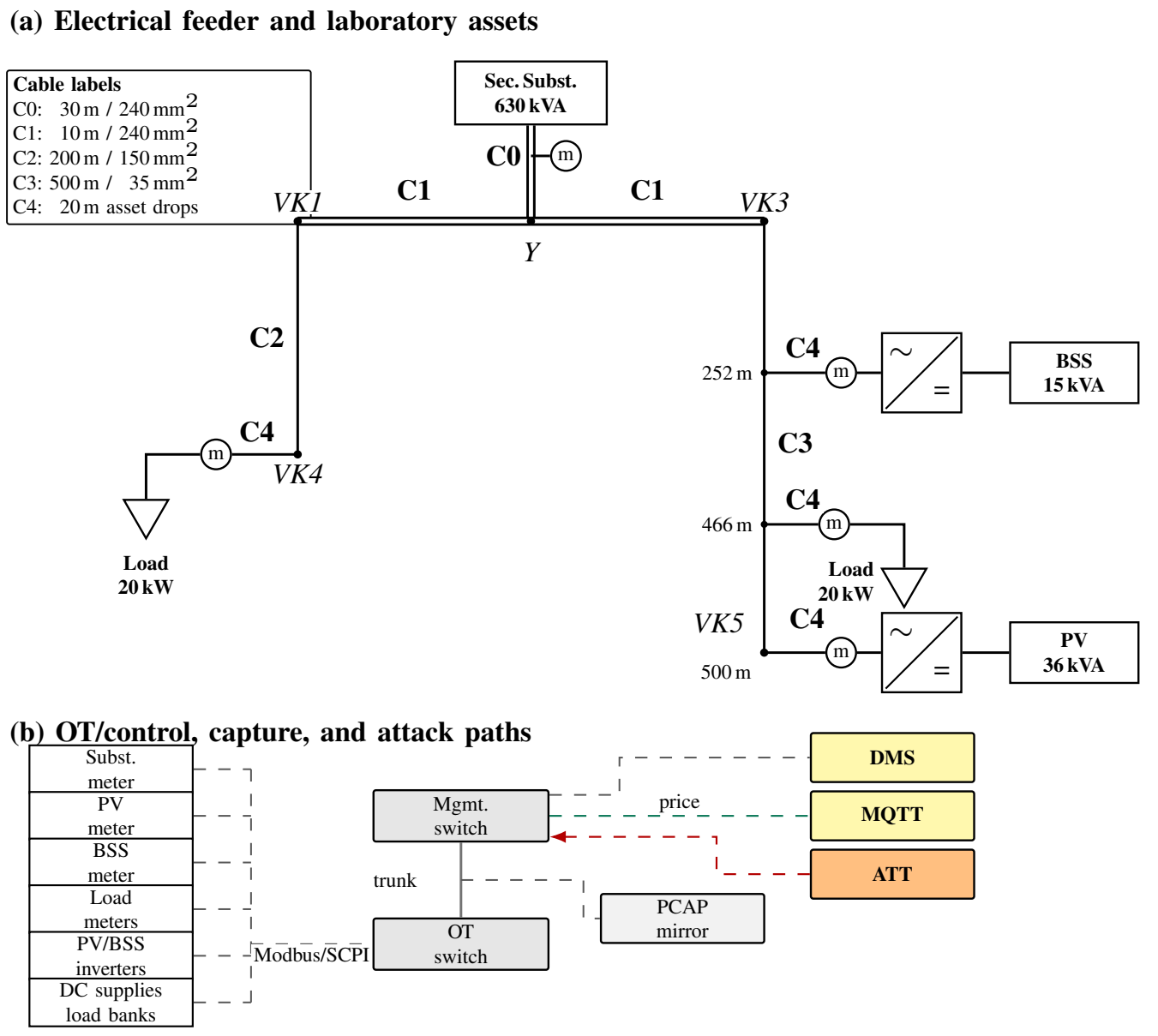
\begin{figure}[t]
\centering
\resizebox{\linewidth}{!}{
\begin{tikzpicture}[
    font=\scriptsize,
    every node/.style={inner sep=0.7pt},
    paneltitle/.style={font=\scriptsize\bfseries, anchor=west},
    bus/.style       ={draw=black, line width=0.62pt},
    parbus/.style    ={draw=black, line width=0.62pt, double, double distance=1.0pt},
    seglabel/.style  ={font=\scriptsize\bfseries, fill=white, inner sep=0.8pt},
    taplabel/.style  ={font=\tiny, fill=white, inner sep=0.7pt, align=right},
    pathlabel/.style ={font=\tiny, fill=white, inner sep=0.6pt},
    junc/.style      ={circle, fill=black, inner sep=0pt, minimum size=2pt},
    tap/.style       ={circle, fill=black, inner sep=0pt, minimum size=1.7pt},
    juncLab/.style   ={font=\scriptsize\itshape, fill=white, inner sep=0.5pt},
    meter/.style     ={circle, draw=black, line width=0.5pt, fill=white,
                       inner sep=0pt, minimum size=2.55mm, font=\tiny},
    inverter/.style  ={draw=black, line width=0.5pt, fill=white,
                       inner sep=0pt, minimum width=6.8mm, minimum height=6.8mm,
                       path picture={
                         \draw[black, line width=0.45pt]
                           ([xshift=0.9pt,yshift=0.9pt]path picture bounding box.south west) --
                           ([xshift=-0.9pt,yshift=-0.9pt]path picture bounding box.north east);
                         \node[font=\scriptsize, inner sep=0pt]
                           at ([xshift=-1.75mm,yshift=1.65mm]path picture bounding box.center) {$\sim$};
                         \node[font=\scriptsize, inner sep=0pt]
                           at ([xshift=1.75mm,yshift=-1.65mm]path picture bounding box.center) {=};
                       }},
    asset/.style     ={draw=black, line width=0.5pt, fill=white,
                       inner sep=0.9pt, align=center, font=\tiny\bfseries,
                       minimum width=11mm, minimum height=5.2mm},
    loadtri/.style   ={draw=black, line width=0.5pt, fill=white, regular polygon,
                       regular polygon sides=3, shape border rotate=180,
                       inner sep=0.4pt, minimum size=4.4mm},
    otasset/.style   ={draw=black, line width=0.45pt, fill=white,
                       inner sep=0.8pt, align=center, font=\tiny,
                       minimum width=14mm, minimum height=4.1mm},
    sw/.style        ={draw=black, line width=0.5pt, rounded corners=0.6pt,
                       fill=black!10, inner sep=0.8pt,
                       minimum width=15mm, minimum height=4.4mm, font=\tiny,
                       align=center},
    dmsbox/.style    ={draw=black, line width=0.5pt, rounded corners=0.6pt,
                       fill=yellow!40, inner sep=0.8pt,
                       minimum width=14mm, minimum height=4.2mm,
                       font=\tiny\bfseries, align=center},
    attbox/.style    ={draw=black, line width=0.5pt, rounded corners=0.6pt,
                       fill=orange!50, inner sep=0.8pt,
                       minimum width=14mm, minimum height=4.2mm,
                       font=\tiny\bfseries, align=center},
    capbox/.style    ={draw=black, line width=0.5pt, rounded corners=0.6pt,
                       fill=black!5, inner sep=0.8pt,
                       minimum width=14mm, minimum height=4.2mm, font=\tiny,
                       align=center},
    legendbox/.style ={draw=black, line width=0.4pt, rounded corners=0.6pt,
                       fill=white, inner sep=1.5pt, font=\tiny, align=left},
    commline/.style  ={draw=black!65, dashed, line width=0.42pt},
    mqttline/.style  ={draw=d4cTeal!80!black, dashed, line width=0.46pt},
    attackline/.style={draw=red!70!black, dashed, line width=0.46pt, -{Latex[length=1.4mm]}},
    x=1mm, y=1mm,
  ]

  \node[paneltitle] at (0,86) {(a) Electrical feeder and laboratory assets};

  \node[legendbox, anchor=north west, text width=25mm] at (0,82) {%
    \textbf{Cable labels}\\
    C0: \phantom{0}30\,m / 240\,mm$^2$\\
    C1: \phantom{0}10\,m / 240\,mm$^2$\\
    C2: 200\,m / 150\,mm$^2$\\
    C3: 500\,m / \phantom{0}35\,mm$^2$\\
    C4: \phantom{0}20\,m asset drops
  };

  \node[asset, minimum width=13mm, minimum height=5.4mm]
       (sub) at (45,80) {Sec.\,Subst.\\\tiny 630\,kVA};
  \coordinate (yJ) at (45,69);
  \draw[parbus] (sub.south) -- (yJ);
  \node[meter] (msub) at (48,74.6) {m};
  \draw[bus] (msub.west) -- (45,74.6);
  \node[seglabel, left=0.5mm] at (45,74.6) {C0};

  \coordinate (vk1) at (25,69);
  \coordinate (vk3) at (65,69);
  \draw[parbus] (vk1) -- (yJ);
  \draw[parbus] (yJ) -- (vk3);

  \node[junc] at (yJ) {};
  \node[juncLab, anchor=north] at ([yshift=-1.6mm]yJ) {Y};
  \node[junc] at (vk1) {};
  \node[junc] at (vk3) {};
  \node[juncLab, above=0.5mm of vk1] {VK1};
  \node[juncLab, above=0.5mm of vk3] {VK3};
  \node[seglabel] at ($(vk1)!0.50!(yJ)+(0,2.7)$) {C1};
  \node[seglabel] at ($(yJ)!0.50!(vk3)+(0,2.7)$) {C1};

  \coordinate (vk4) at (25,49);
  \draw[bus] (vk1) -- (vk4);
  \node[junc] at (vk4) {};
  \node[juncLab, below=0.6mm of vk4] {VK4};
  \node[seglabel, left=0.8mm] at (25,59) {C2};

  \node[meter] (mLoadL) at (18,49) {m};
  \draw[bus] (vk4) -- (mLoadL.east);
  \node[loadtri] (loadL) at (12,44.0) {};
  \draw[bus] (mLoadL.west) -- (12,49) -- (loadL.north);
  \node[font=\tiny\bfseries, anchor=north, align=center]
       at ([yshift=-1mm]loadL.south) {Load\\20\,kW};
  \node[seglabel, above=0.7mm] at (21.5,49) {C4};

  \coordinate (tapBSS) at (65,56);
  \coordinate (tapLoad) at (65,43);
  \coordinate (vk5) at (65,32);
  \draw[bus] (vk3) -- (vk5);
  \node[tap] at (tapBSS) {};
  \node[tap] at (tapLoad) {};
  \node[junc] at (vk5) {};
  \node[juncLab, anchor=south east, inner sep=1pt]
       at ([xshift=-1.3mm, yshift=1.3mm]vk5) {VK5};
  \node[seglabel, right=0.8mm] at (65,50) {C3};
  \node[taplabel, left=0.7mm] at (tapBSS) {252\,m};
  \node[taplabel, left=0.7mm] at (tapLoad) {466\,m};
  \node[taplabel, anchor=north east] at ([xshift=-0.8mm,yshift=-0.7mm]vk5) {500\,m};

  \node[meter] (mBSS) at (71.6,56) {m};
  \node[inverter] (invBSS) at (78.5,56) {};
  \node[asset, anchor=west] (bss) at (86,56) {BSS\\\tiny 15\,kVA};
  \draw[bus] (tapBSS) -- (mBSS.west);
  \draw[bus] (mBSS.east) -- (invBSS.west);
  \draw[bus] (invBSS.east) -- (bss.west);
  \node[seglabel, above=0.9mm] at (68.3,56) {C4};

  \node[meter] (mLoadR) at (71.0,43) {m};
  \draw[bus] (tapLoad) -- (mLoadR.west);
  \node[loadtri] (loadR) at (77,38.1) {};
  \draw[bus] (mLoadR.east) -- (77,43) -- (loadR.north);
  \node[font=\tiny\bfseries, anchor=east, align=right, fill=white, inner sep=0.4pt]
       at ([xshift=-1.0mm]loadR.west) {Load\\20\,kW};
  \node[seglabel, above=0.9mm] at (68.3,43) {C4};

  \node[meter] (mPV) at (71.6,32) {m};
  \node[inverter] (invPV) at (78.5,32) {};
  \node[asset, anchor=west] (pv) at (86,32) {PV\\\tiny 36\,kVA};
  \draw[bus] (vk5) -- (mPV.west);
  \draw[bus] (mPV.east) -- (invPV.west);
  \draw[bus] (invPV.east) -- (pv.west);
  \node[seglabel, anchor=south west] at (66.8,33.7) {C4};

  \node[paneltitle] at (0,25) {(b) OT/control, capture, and attack paths};

  \node[otasset] (fd1) at (9,22) {Subst.\\meter};
  \node[otasset] (fd2) at (9,18) {PV\\meter};
  \node[otasset] (fd3) at (9,14) {BSS\\meter};
  \node[otasset] (fd4) at (9,10) {Load\\meters};
  \node[otasset] (fd5) at (9,6) {PV/BSS\\inverters};
  \node[otasset] (fd6) at (9,2) {DC supplies\\load banks};

  \node[sw]     (ot)   at (39,7)  {OT\\switch};
  \node[sw]     (mgmt) at (39,18) {Mgmt.\\switch};
  \node[capbox] (tap)  at (58,9.2) {PCAP\\mirror};
  \node[dmsbox] (dms)  at (76,23) {DMS};
  \node[dmsbox] (mqtt) at (76,18) {MQTT};
  \node[attbox] (att)  at (76,13) {ATT};

  \coordinate (fieldbusTop) at (21,22);
  \coordinate (fieldbusBot) at (21,2);
  \draw[commline] (fieldbusTop) -- (fieldbusBot);
  \draw[commline] (fd1.east) -- (21,22);
  \draw[commline] (fd2.east) -- (21,18);
  \draw[commline] (fd3.east) -- (21,14);
  \draw[commline] (fd4.east) -- (21,10);
  \draw[commline] (fd5.east) -- (21,6);
  \draw[commline] (fd6.east) -- (21,2);
  \draw[commline] (21,7) -- node[pathlabel, below] {Modbus/SCPI} (ot.west);

  \draw[bus, draw=black!55] (ot.north) -- (mgmt.south);
  \node[pathlabel, anchor=east] at (35.5,12.5) {trunk};
  \draw[commline] (39,12.5) -- (49.5,12.5) |- (tap.west);

  \draw[commline] ([yshift=1.8mm]mgmt.east) -- ++(6,0) |- (dms.west);
  \draw[mqttline] (mgmt.east) -- node[pathlabel, above] {price} (mqtt.west);
  \draw[attackline] (att.west) -- ++(-8,0) |- ([yshift=-1.8mm]mgmt.east);

\end{tikzpicture}}
\caption{Laboratory topology. Panel (a) shows the electrical feeder with metering points and DER/load assets along the 500\,m VK3--VK5 tapped line (BSS at 252\,m, second load bank at 466\,m, PV at VK5). Panel (b) separates the field-level OT switch (meters, inverters, DC supplies, load banks) from the management switch (DMS, MQTT broker, attacker host); a passive mirror/tap on the inter-switch trunk records OT traffic.}
\label{fig:lab_detailed}
\end{figure}

\section{Methodology}
The methodology translates the literature gap into an experimental data-generation process. The design choices are driven by three constraints: cyber events must be observable at the network layer, their effects must be independently measurable in the physical feeder, and the resulting samples must preserve temporal order for stateful IDS evaluation.

\subsection{Use Case and Requirements}
We target a distribution-management scenario for behind-the-meter and feeder-connected assets that performs self-consumption optimization and price-aware dispatch consistent with established multi-use behind-the-meter flexibility studies~\cite{r24}. The requirements below operationalize the gaps identified in Sec.~II: prior datasets rarely combine real DER hardware, mixed OT protocols, control context, and phase-resolved labels in a form suitable for temporally ordered IDS training.
\begin{itemize}
  \item \textbf{Cross-domain observability:} aligned network traces, controller context, and independent physical measurements.
  \item \textbf{Temporal validity:} consistent timestamps across sources to correlate cyber actions and physical effects without learning alignment artifacts.
  \item \textbf{Attack diversity:} multi-stage campaigns covering protocol-semantic, telemetry, and context-channel manipulation.
  \item \textbf{Label fidelity:} phase-resolved labels with explicit start/end times and interpretable descriptions.
  \item \textbf{Reusable representation:} machine-readable tables, packet traces, state views, and scenario metadata that support independent IDS studies.
\end{itemize}

\subsection{Testbed and Instrumentation}
\begin{figure}[!t]
\centering
\resizebox{\linewidth}{!}{
\begin{tikzpicture}[node distance=5mm, font=\footnotesize]

  \node[d4cBox, text width=30mm] (grid) {%
    \textbf{Electrical layer}\\[-0.2em]
    Substation meter\\
    PV inverter\\
    Battery inverter\\
    Controllable loads
  };

  \node[d4cBox, text width=30mm, right=20mm of grid] (ot) {%
    \textbf{OT network}\\[-0.2em]
    Modbus/TCP polling\\
    Modbus/TCP writes\\
    MQTT price channel
  };

  \node[d4cVM, text width=25mm, above=8mm of ot] (ems) {%
    DMS/EMS\\controller
  };

  \node[d4cBox, text width=30mm, right=20mm of ot] (data) {%
    \textbf{Data capture}\\[-0.2em]
    Passive PCAP tap\\
    Time-series export\\
    Attack/controller logs
  };

  \node[d4cAttacker, text width=24mm, below=8mm of ot] (adv) {%
    Adversary\\tooling
  };

  \draw[d4cArrowTeal] (grid.east) -- node[above,d4cLabel] {measurements} (ot.west);
  \draw[d4cArrowBlue] (ot.north) -- node[right,d4cLabel] {telemetry / setpoints} (ems.south);
  \draw[d4cArrow] (ot.east) -- node[above,d4cLabel] {mirrored traffic} (data.west);
  \draw[d4cArrowRedDashed] (adv.north) -- node[right,d4cLabel] {recon / manipulation} (ot.south);
  \draw[d4cArrowRedDashed] (adv.west) to[bend right=14] node[below,d4cLabel] {process impact} (grid.south);

\end{tikzpicture}}
\caption{Layered abstraction of the laboratory testbed used throughout the paper. The electrical layer feeds measurements via the OT network; mirrored OT traffic and parallel telemetry exports flow to the data-capture layer; the DMS/EMS controller exchanges Modbus/TCP polling, Modbus/TCP writes, and MQTT price updates over the OT network; adversary-controlled paths (dashed red) target the OT network for reconnaissance, manipulation, and process impact.}
\label{fig:lab_environment}
\end{figure}
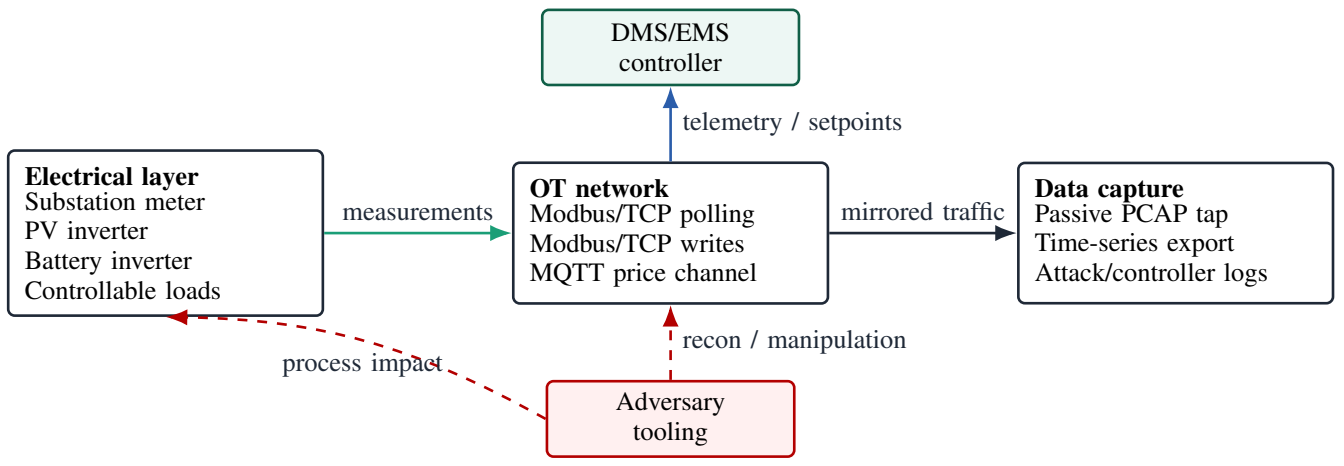

The testbed topology in Fig.~\ref{fig:lab_detailed} is designed to create measurable feeder-level consequences from localized DER manipulation; Fig.~\ref{fig:lab_environment} abstracts the same setup into the layered electrical/OT/data-capture view used throughout the paper. The electrical feeder starts at a 630~kVA secondary substation and reaches distribution node Y through a 30~m, 240~mm$^2$ cable monitored by a three-phase meter. Y branches through two 10~m, 240~mm$^2$ cables to VK1 and VK3. The VK1 branch continues over 200~m with 150~mm$^2$ cross-section to VK4, where a 20~m drop connects a controllable 20~kW load bank with AC-side three-phase metering. The VK3 branch is a 500~m, 35~mm$^2$ feeder to VK5: a 252~m tap connects a 15~kVA BSS through a 20~m drop, a 466~m tap connects the second 20~kW load bank through a 20~m drop, and VK5 connects a 36~kVA PV inverter through a 20~m drop. These cable lengths, cross-sections, and asset ratings create realistic impedance separation between substation, storage, load, and PV nodes, so that local manipulation can propagate into observable active-power and storage-state deviations rather than remaining a purely logical network event. All AC-side meters are attached to the OT switch and provide Modbus/TCP telemetry~\cite{r7}; PV/BSS inverters receive Modbus setpoints through the same OT network. The DC supplies behind the inverters and the load banks are controllable (SCPI for DC sources, Modbus for loads), which allows PV and load time series to be replayed physically during experiments. A management switch connects the DMS, MQTT price-signal server~\cite{r8}, and attacker host with physical network access. A passive mirror/tap records the inter-switch trunk.

\subsection{DMS Operating Logic: VED EMS for Self-Consumption Optimization}
The DMS hosts a supervisory energy management system (EMS) running on a virtual edge device (VED) that dispatches DER setpoints at fixed time steps. In the hierarchy used here, the VED is the local execution platform, the EMS implements the dispatch policy, and the DMS coordinates wide-area context such as MQTT price signals. The baseline objective is self-consumption optimization (SCO), maximizing local PV usage while respecting battery power and state-of-charge (SoC) constraints. An external price incentive can shift charging/discharging decisions when operational constraints are met. Algorithm~\ref{alg:sco} summarizes the dispatch logic.

\ifdefined\algorithm
\begin{algorithm}[t]
\caption{VED EMS dispatch logic for SCO with price signal integration (SCO+).}
\label{alg:sco}
\footnotesize
\begin{algorithmic}[1]
\STATE \textbf{Inputs:} measured $P_{\mathrm{pv}}(t)$, $P_{\mathrm{load}}(t)$, $P_{\mathrm{bat}}(t)$; price $\pi(t)$; last $P_{\mathrm{bat}}(t-\Delta t)$; SOC estimate $\widehat{\mathrm{SOC}}(t-\Delta t)$; battery limits $P_{\mathrm{bat}}^{\min},\,P_{\mathrm{bat}}^{\max}$; SOC band $\mathrm{SOC}_{\min},\,\mathrm{SOC}_{\max}$; blending factor $\alpha\in[0,1]$; correction gain $\kappa$.
\STATE Compute net household power without battery: $P_{\mathrm{house}}(t)\leftarrow P_{\mathrm{load}}(t)+P_{\mathrm{pv}}(t)$.
\STATE Update SOC estimate by trapezoidal integration.
\STATE If reference SOC measurement $\mathrm{SOC}_{\mathrm{ref}}(t)$ exists, apply observer correction.
\STATE Baseline SCO setpoint: $P_{\mathrm{sco}}(t)\leftarrow -P_{\mathrm{house}}(t)$.
\IF{$\pi(t)$ is available}
\STATE Normalize price and compute price-driven setpoint.
\STATE Blend strategies: $P_{\mathrm{cmd}}(t)\leftarrow \alpha\cdot P_{\mathrm{price}}(t)+(1-\alpha)\cdot P_{\mathrm{sco}}(t)$.
\ELSE
\STATE $P_{\mathrm{cmd}}(t)\leftarrow P_{\mathrm{sco}}(t)$.
\ENDIF
\STATE Enforce SOC protection band and clamp to battery limits.
\STATE \textbf{Output:} dispatch $P^*_{\mathrm{bat}}(t)$ to the battery interface.
\end{algorithmic}
\end{algorithm}
\fi
The blending step in Algorithm~\ref{alg:sco} is also a deliberate attack surface. Because the command $P_{\mathrm{cmd}}(t)$ mixes the physical self-consumption objective with an external price-driven setpoint, manipulation of the MQTT price channel changes BSS dispatch without modifying Modbus polling or inverter registers directly. This coupling motivates the S6 scenario and makes single-channel network thresholds insufficient.

\subsection{Acquisition, Time Alignment, and Processing}
\begin{figure}[t]
\centering
\begin{tikzpicture}[node distance=4mm, font=\small]

  \node[d4cBox, text width=0.94\linewidth] (inputs) {%
    \textbf{Inputs}\\[-0.2em]
    \textbullet\ Time-series telemetry exports and controller context\\
    \textbullet\ Mirrored OT traffic captures (Modbus/TCP, MQTT)\\
    \textbullet\ Attacker/controller traces when available
  };

  \node[d4cBox, text width=0.94\linewidth, below=of inputs] (proc) {%
    \textbf{Processing (default step size: 1\,s)}\\[-0.2em]
    \textbullet\ Normalize names, units, and timestamps\\
    \textbullet\ Align all sources on a fixed 1~s grid\\
    \textbullet\ Build fixed-step feature tables and phase labels\\
    \textbullet\ Prepare full/benign/attack PCAP views aligned with the run timeline\\
    \textbullet\ Preserve scenario and testbed context in metadata
  };

  \node[d4cBox, text width=0.94\linewidth, below=of proc] (out) {%
    \textbf{Synchronized data model}\\[-0.2em]
    \textbullet\ Fixed-step tables: main + benign/attack splits\\
    \textbullet\ PCAP views: full + benign/attack slices\\
    \textbullet\ Semantics: phase and scenario definitions\\
    \textbullet\ Metadata: experiment and testbed infrastructure descriptors\\
    \textbullet\ State view with attack-phase context for replay/stateful IDS
  };

  \draw[d4cArrow] (inputs.south) -- (proc.north);
  \draw[d4cArrow] (proc.south) -- (out.north);

\end{tikzpicture}
\caption{Dataset-generation pipeline. Raw telemetry, OT packet captures, and controller logs are synchronized to a fixed step size and mapped into a multi-layer data model with phase-resolved labels, scenario context, and testbed metadata.}
\label{fig:dataset_pipeline}
\end{figure}
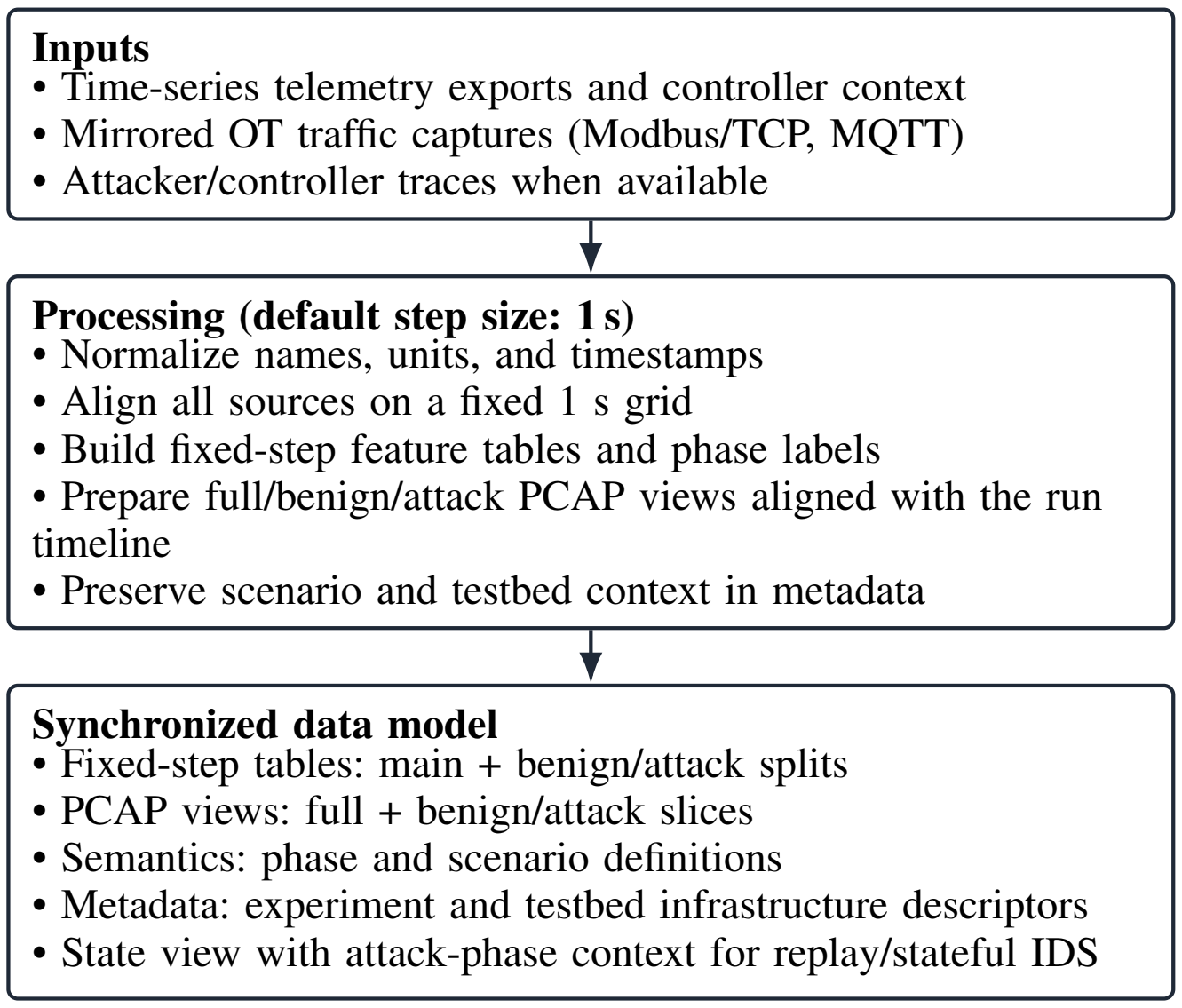

Fig.~\ref{fig:dataset_pipeline} summarizes acquisition and post-processing. Three acquisition paths run in parallel during a scenario: physical signals are sampled at the meters with a 1\,s SCPI cadence, the OT switch trunk is mirrored to a passive PCAP recorder, and the EMS controller logs its own setpoints, observer states, and attacker-injected events. The 1\,s alignment is a methodological control: it reduces temporal jitter between Modbus packets, controller decisions, and physical measurements, preventing downstream models from learning synchronization artifacts instead of process behavior. Per-step labels are materialized as a binary attack indicator, a primary phase label, and an all-phases label that retains overlapping phase membership. Each scenario is represented by synchronized tables, PCAP views, phase intervals, device/register mappings, and scenario metadata, consistent with FAIR-oriented design principles~\cite{r22}. Sensitive identifiers are normalized to laboratory-only values to avoid operational disclosure concerns.

\section{Experiments}
The experiments instantiate the methodological requirements as controlled but physically grounded DMS runs. One benign baseline establishes normal SCO/SCO+ behavior, while six attack scenarios vary the manipulated channel, target asset, and expected process observability.

\subsection{Scenarios and Assumptions}
Experiments are executed as bounded 30-minute runs under controlled load situations and predefined PV/load profiles. This duration is long enough to expose attack onset, process propagation, and recovery behavior while keeping cause-effect attribution traceable. The supervisory controller performs SCO/SCO+ dispatch as described in Sec.~III and Algorithm~\ref{alg:sco}, independent meters provide physical ground truth, and network traffic is captured passively. For S1, one representative PV-manipulation run is used in aggregate metrics to avoid over-weighting one attack family.

\subsection{Attack Catalog Instantiation}
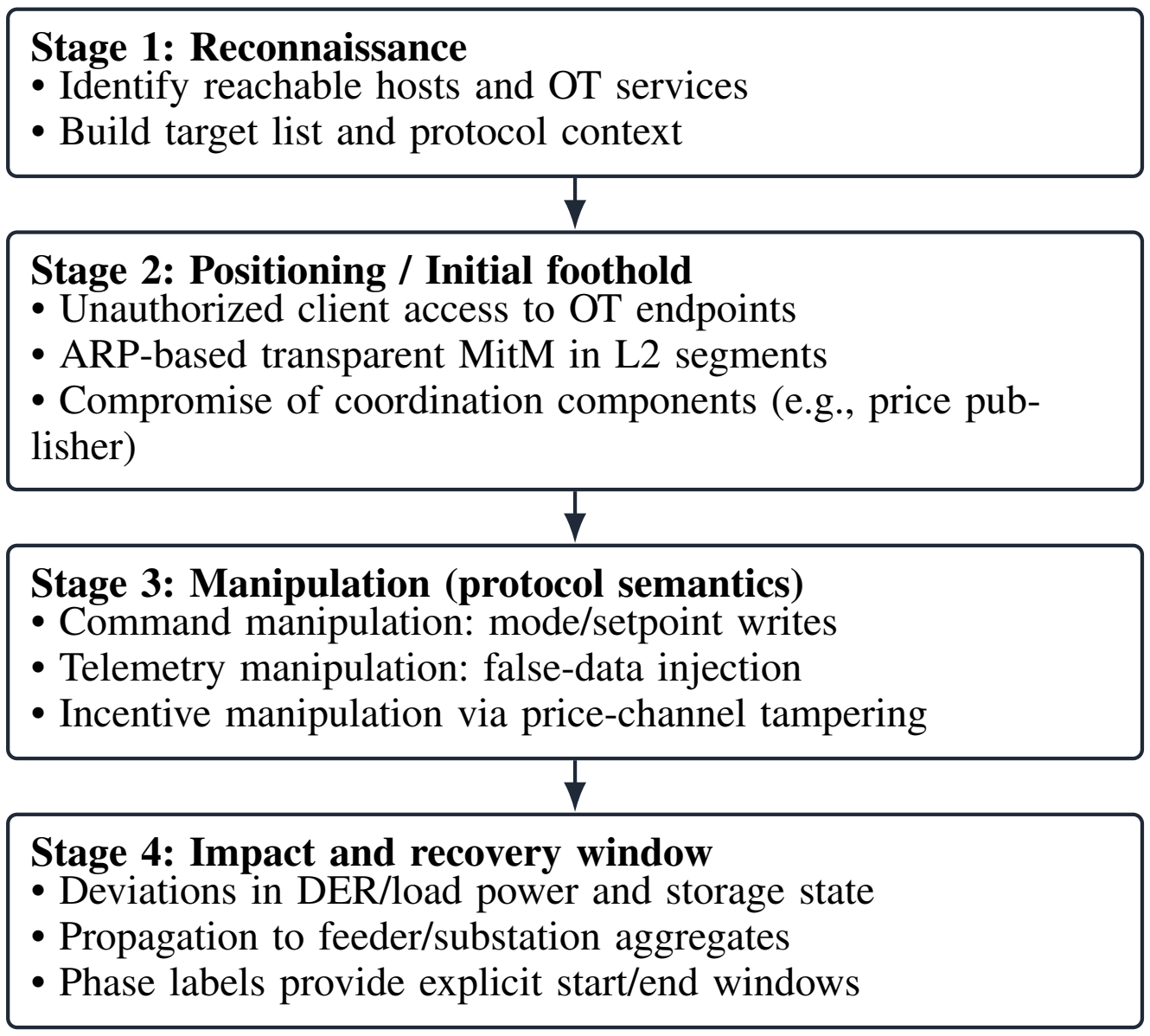
\begin{figure}[t]
\centering
\begin{tikzpicture}[node distance=4mm, font=\small]

  \node[d4cBox, text width=0.94\linewidth] (s1) {%
    \textbf{Stage 1: Reconnaissance}\\[-0.2em]
    \textbullet\ Identify reachable hosts and OT services\\
    \textbullet\ Build target list and protocol context
  };

  \node[d4cBox, text width=0.94\linewidth, below=of s1] (s2) {%
    \textbf{Stage 2: Positioning / Initial foothold}\\[-0.2em]
    \textbullet\ Unauthorized client access to OT endpoints\\
    \textbullet\ ARP-based transparent MitM in L2 segments\\
    \textbullet\ Compromise of coordination components (e.g., price publisher)
  };

  \node[d4cBox, text width=0.94\linewidth, below=of s2] (s3) {%
    \textbf{Stage 3: Manipulation (protocol semantics)}\\[-0.2em]
    \textbullet\ Command manipulation: mode/setpoint writes\\
    \textbullet\ Telemetry manipulation: false-data injection\\
    \textbullet\ Incentive manipulation via price-channel tampering
  };

  \node[d4cBox, text width=0.94\linewidth, below=of s3] (s4) {%
    \textbf{Stage 4: Impact and recovery window}\\[-0.2em]
    \textbullet\ Deviations in DER/load power and storage state\\
    \textbullet\ Propagation to feeder/substation aggregates\\
    \textbullet\ Phase labels provide explicit start/end windows
  };

  \draw[d4cArrow] (s1.south) -- (s2.north);
  \draw[d4cArrow] (s2.south) -- (s3.north);
  \draw[d4cArrow] (s3.south) -- (s4.north);

\end{tikzpicture}
\caption{Attack procedure overview (reconnaissance to impact) used as a template for phase-resolved labeling.}
\label{fig:attack_procedure}
\end{figure}

We instantiate representative attack patterns against DER management and its communication. Fig.~\ref{fig:attack_procedure} provides the conceptual sequence from reconnaissance to impact that is reflected in phase labels, while Table~\ref{tab:runs} maps each scenario to its manipulated channel and expected process impact.
\begin{itemize}
  \item \textbf{Reconnaissance:} host/service discovery to identify Modbus/TCP targets.
  \item \textbf{Protocol-semantic manipulation (Industroyer-inspired):} unauthorized Modbus client actions modifying setpoints or operating modes (S1, S2).
  \item \textbf{Transparent manipulation (MitM/FDI):} ARP-spoofing-based telemetry manipulation in meter-to-DMS paths (S3, S4, S5).
  \item \textbf{Price-signal manipulation:} compromised MQTT price publication path to bias EMS dispatch decisions (S6).
\end{itemize}

\paragraph{Relationship to IT-style attack stages}
The scenarios isolate OT-resident execution and impact phases to keep cyber-physical cause-effect labels clean. Their prerequisites still map onto well-known MITRE ATT\&CK for ICS techniques~\cite{r26}: \textit{network service scanning} and \textit{remote system discovery} (S1--S5 reconnaissance), \textit{adversary-in-the-middle} via \textit{ARP cache poisoning} (S3--S5), \textit{unauthorized command message} via a credentialed Modbus client (S1, S2), and \textit{supply-chain compromise} of an upstream price-signal publisher (S6). IT-stage precursors such as credential theft, phishing, lateral movement, or ransomware delivery are treated as assumed upstream conditions rather than measured phases, which prevents unrelated IT noise from diluting the process labels.

\begin{table}[!t]
\caption{Primary scenarios, attack families, and impact observations.}
\label{tab:runs}
\centering
\footnotesize
\setlength{\tabcolsep}{4pt}
\renewcommand{\arraystretch}{1.12}
\begin{tabular}{L{0.55cm} L{1.45cm} L{2.05cm} L{2.95cm}}
\toprule
\textbf{ID} & \textbf{Family} & \textbf{Primary Target/Effect} & \textbf{Impact Observation}\\
\midrule
S0 & Benign & Normal EMS operation. & Baseline trajectories without malicious perturbation.\\
S1 & Modbus semantic manipulation & PV control behavior. & PV-side active-power response deviates from expected supervisory profile.\\
S2 & Modbus semantic manipulation & BSS dispatch behavior. & BSS active-power trajectory shows unauthorized dispatch behavior.\\
S3 & ARP spoofing/FDI & BSS meter path; SoC bias. & Reported BSS-related values are biased, shifting DMS SoC-relevant inference.\\
S4 & Two-phase ARP spoofing/FDI & Load/PV paths; forced dispatch inversion. & Residual-load interpretation flips across phases, inducing opposite charge/discharge decisions.\\
S5 & Two-phase ARP spoofing/FDI & Load/PV/BSS paths; SoC divergence. & Divergence grows between reference and reported SoC during phase 2.\\
S6 & MQTT supply-chain compromise & Price-signal manipulation; stealthy context path. & Dispatch changes follow manipulated incentives with limited lower-layer Modbus evidence.\\
\bottomrule
\end{tabular}
\end{table}

\subsection{Experiment Characterization and Visualizations}
\begin{figure}[!t]
\centering
\includegraphics[width=\linewidth]{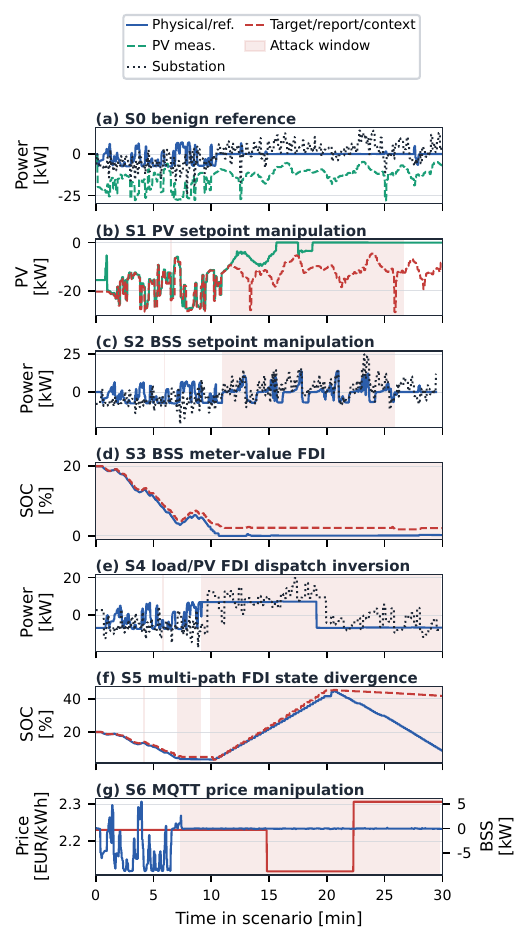}
\caption{Scenario-complete cyber-physical comparison. The subplots show the most diagnostic signal family for S0--S6: benign power variation, PV profile mismatch, BSS setpoint manipulation, BSS meter-value FDI, load/PV FDI, multi-path SoC divergence, and MQTT price manipulation. Blue marks physical/reference signals, red marks target/reported/context variables, and shaded regions mark attack intervals.}
\label{fig:results_comparison}
\end{figure}

Phase intervals map each run to per-step attack indicators and categorical phase labels. Fig.~\ref{fig:results_comparison} compares the benign reference with the most diagnostic signal family from every primary attack scenario. S0 provides the normal SCO/SCO+ power reference. S1 highlights PV-side active-power deviation from the expected profile, while S2 shows direct BSS dispatch manipulation. S3 and S5 expose reported-versus-reference SoC divergence under FDI, with S5 producing the stronger multi-path inconsistency. S4 shows dispatch inversion when load and PV measurements are falsified together. S6 demonstrates a context-channel attack, where manipulated price information changes BSS behavior without requiring a lower-layer Modbus FDI signature.

\section{Dataset}
\label{sec:dataset}

This section translates the experimental campaign into the reusable dataset structure. It first explains the scenario organization and synchronized signal layers, then separates variable semantics from data formats, label representation, and data-quality controls. The aim is to clarify which information is required for scientific reuse, which channels should be excluded from detector inputs to avoid feature leakage, and how the dataset supports both process-level analysis and network-oriented IDS studies.

\subsection{Suite Overview}
Data4Cyber is organized by scenario so that each run preserves its own timing, labels, network traces, and process context. The benign S0 scenario supports calibration and false-positive analysis, while S1--S6 provide attack diversity across direct command manipulation, transparent telemetry falsification, and context-channel compromise. The primary set comprises 12{,}562 synchronized 1\,s state messages (4{,}998 benign, 7{,}564 attack), enough volume for both unsupervised calibration on S0 and chronologically blocked supervised evaluation across the attack scenarios.

\subsection{Data Sources and Signals}
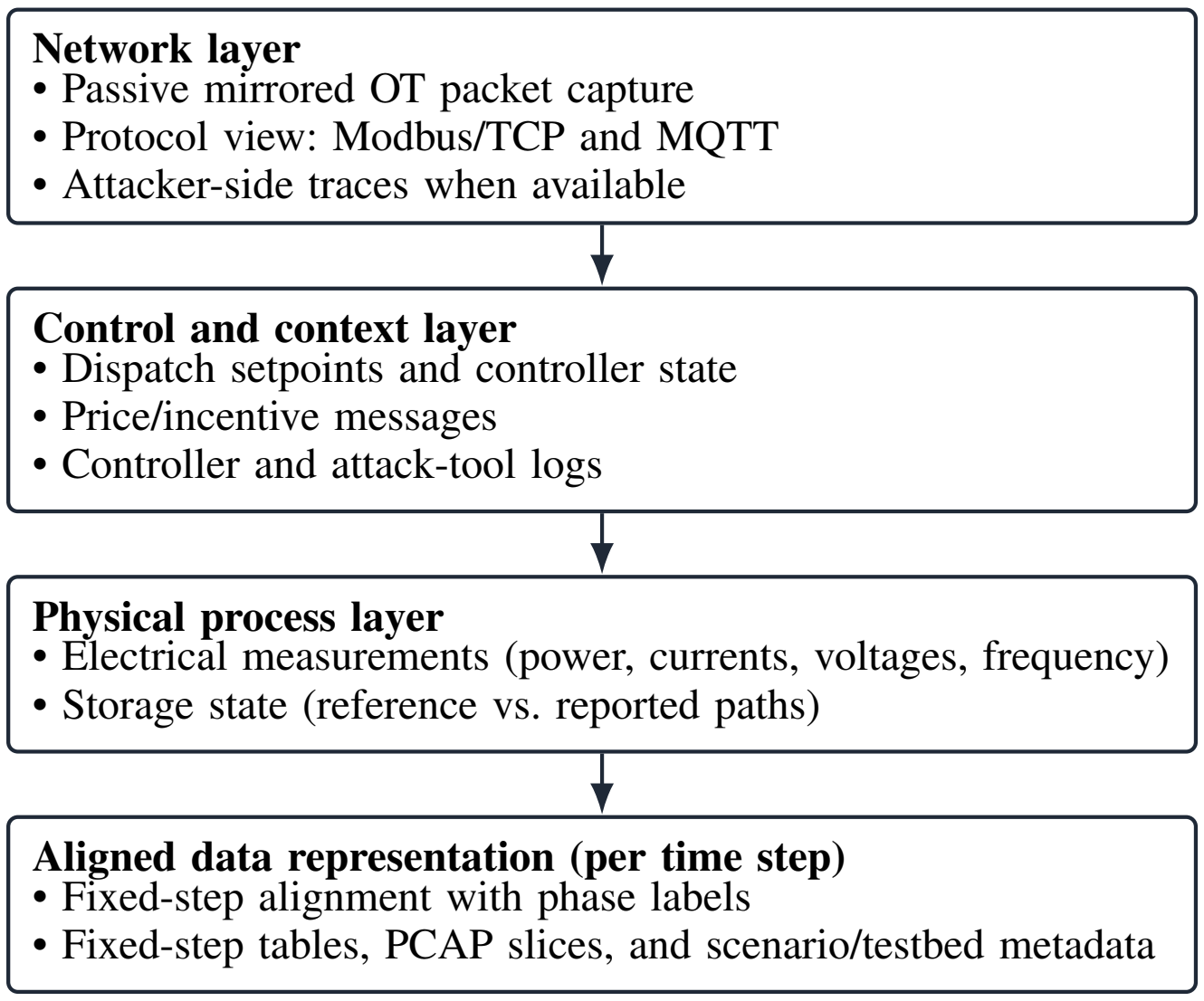
\begin{figure}[t]
\centering
\begin{tikzpicture}[node distance=4.5mm, font=\small]

  \node[d4cBox, text width=0.94\linewidth] (net) {%
    \textbf{Network layer}\\[-0.2em]
    \textbullet\ Passive mirrored OT packet capture\\
    \textbullet\ Protocol view: Modbus/TCP and MQTT\\
    \textbullet\ Attacker-side traces when available
  };

  \node[d4cBox, text width=0.94\linewidth, below=of net] (ctl) {%
    \textbf{Control and context layer}\\[-0.2em]
    \textbullet\ Dispatch setpoints and controller state\\
    \textbullet\ Price/incentive messages\\
    \textbullet\ Controller and attack-tool logs
  };

  \node[d4cBox, text width=0.94\linewidth, below=of ctl] (phy) {%
    \textbf{Physical process layer}\\[-0.2em]
    \textbullet\ Electrical measurements (power, currents, voltages, frequency)\\
    \textbullet\ Storage state (reference vs. reported paths)
  };

  \node[d4cBox, text width=0.94\linewidth, below=of phy] (rel) {%
    \textbf{Aligned data representation (per time step)}\\[-0.2em]
    \textbullet\ Fixed-step alignment with phase labels\\
    \textbullet\ Fixed-step tables, PCAP slices, and scenario/testbed metadata
  };

  \draw[d4cArrow] (net.south) -- (ctl.north);
  \draw[d4cArrow] (ctl.south) -- (phy.north);
  \draw[d4cArrow] (phy.south) -- (rel.north);

\end{tikzpicture}
\caption{Multi-layer observability in Data4Cyber. Three synchronized observability layers (network, control/context, physical) feed an aligned per-step data representation that supports network IDS research, DMS decision analysis, and cyber-physical impact validation.}
\label{fig:data_sources}
\end{figure}

Fig.~\ref{fig:data_sources} summarizes the multi-layer observability. Four sources are integrated on the same 1\,s grid: AC-meter telemetry from the substation, PV inverter, BSS inverter, and the two controllable loads provides three-phase active power, line currents, line-to-neutral voltages, frequency, and power factors; the MQTT price channel provides the EMS incentive signal; the battery state of charge is reported through two paths (a reference value read from the BSS itself and a reported value computed by the EMS observer), whose divergence is a primary indicator for FDI scenarios; and mirrored network telemetry captured at the OT switch trunk closes the loop with packet-level evidence.

\begin{table}[!t]
\caption{Main feature groups in the synchronized CSV and state-log files.}
\label{tab:feature_groups}
\centering
\footnotesize
\setlength{\tabcolsep}{3pt}
\renewcommand{\arraystretch}{1.08}
\begin{tabular}{L{2.05cm} L{2.45cm} L{2.7cm}}
\toprule
\textbf{Group} & \textbf{Example Columns} & \textbf{Purpose}\\
\midrule
Power flow & \icode{p_sum3}, \icode{i_sum3}, voltages, frequency & Detect physical deviations and dispatch effects.\\
Storage state & Reference SoC, reported SoC & Analyze reported-vs-reference consistency and FDI impact.\\
Context channel & MQTT price value & Study incentive manipulation and EMS response.\\
Attack observables & Attacker events and old/new values & Link attacker activity to labels where traces exist.\\
Labels & \icode{attack_active}, \icode{attack_phase}, all-phase label & Train/test binary and phase-aware IDS pipelines.\\
\bottomrule
\end{tabular}
\end{table}

\subsection{Data Model and Formats}
The data model is organized into synchronized physical, control/context, network, and label layers. Table~\ref{tab:feature_groups} lists the main signal groups, while each scenario directory provides the following machine-readable views:
\begin{itemize}
  \item \textbf{Fixed-step tables:} CSV tables (main, benign split, attack split) with phase labels.
  \item \textbf{Labels and context:} phase boundaries and scenario semantics in JSON.
  \item \textbf{Network traces:} passive OT PCAP plus benign/attack slices.
  \item \textbf{Mappings:} semantic address/register mappings to device roles and variables.
  \item \textbf{Metadata:} standalone scenario, experiment, and testbed infrastructure descriptors.
  \item \textbf{State view:} per-step state log with attack-phase context for replay-oriented and stateful IDS methods.
\end{itemize}

\subsection{Labels and State Representation}
Each time step carries a binary attack indicator and a categorical phase label that distinguishes reconnaissance, field-value manipulation, false-data injection, and price manipulation. Phase boundaries are derived from attacker and controller traces together with process and packet observations. The state view stores one JSON record per timestamp with the timestamp, the binary attack flag, the phase label, and a flat dictionary of feature values keyed by device and attribute. This representation enables stateful detectors to select semantically meaningful channels, such as PV-inverter active power, BSS state of charge, or MQTT price, without parsing vendor-specific packets.

\subsection{Data Quality and Bias Controls}
\begin{figure}[!t]
\centering
\includegraphics[width=\linewidth]{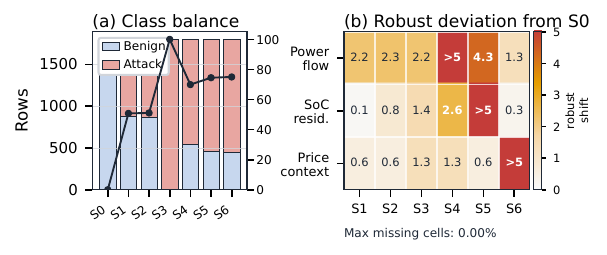}
\caption{Data-quality profile. (a) Benign/attack balance and attack share. (b) Robust scenario observability relative to the benign S0 calibration baseline; values are 90th-percentile deviations normalized by the S0 interquartile range and capped for readability.}
\label{fig:data_quality}
\end{figure}

Data4Cyber is assessed along four data-quality dimensions, summarized in Fig.~\ref{fig:data_quality}. \emph{Temporal consistency} is controlled through strict 1\,s monotonic timestamps and run-bounded PCAP slices, reducing the risk that stateful detectors learn timing artifacts rather than process behavior. \emph{Semantic integrity} is controlled by deriving labels from attacker/controller traces and cross-checking them against process and packet observations, so that phase labels correspond to observable control decisions and physical effects. \emph{Scenario diversity} is introduced through three attack families with different observability profiles: direct Modbus manipulation, transparent FDI on measurement paths, and MQTT context manipulation. \emph{Bias control} is handled by separating benign S0 calibration from attack scenarios, quantifying which feature families deviate from benign calibration, excluding label and attacker-only columns from IDS feature sets, and using chronological or scenario-wise splits for downstream evaluation. The controlled 30-minute laboratory runs intentionally favor traceable cause-effect attribution; broader diurnal, seasonal, and multi-feeder variation remains an extension target rather than a claimed coverage dimension.

\subsection{Baseline IDS Evaluation}
\label{sec:ids_eval}
The open-source IPAL framework~\cite{r21,r25} serves as a standardized baseline harness to characterize dataset difficulty. Because the state logs follow the IPAL schema, compatible IDSs run without custom adapters; the goal is to expose detection difficulty, channel observability, and evaluation pitfalls rather than to propose new IDS designs.

A fixed eight-feature subset is used: active power of the BSS, PV, Load~A, Load~B, and substation, together with the reference SoC, reported SoC, and MQTT price. Label columns and attacker-only observables are excluded to avoid feature leakage. The baseline set contains statistical state detectors, one process-aware detector, and supervised classifiers. The S0-trained detectors are calibrated on benign data and evaluated on S0--S6; supervised classifiers use a chronologically blocked hold-out split, so adjacent samples are not shared between training and test data.

\begin{figure}[!t]
\centering
\includegraphics[width=\linewidth]{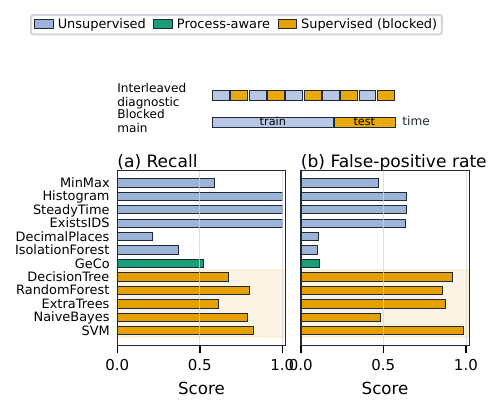}
\caption{Recall/FPR trade-off for twelve baseline IDSs on S0--S6. The inset contrasts the interleaved diagnostic split with the blocked split used for supervised bars, exposing alarm saturation and the lower-FPR region of the process-aware baseline.}
\label{fig:ids_benchmark}
\end{figure}

Table~\ref{tab:ipal_eval} and Fig.~\ref{fig:ids_benchmark} pair recall with false-positive rate. Under the blocked supervised split, F1 ranges from 0.649 to 0.813, with high false-positive rates. The S0-trained baselines show complementary behavior: existence- and histogram-style detectors achieve almost perfect recall at high false-positive rate, whereas DecimalPlaces, Isolation~Forest, and GeCo keep lower false-positive rates with reduced sensitivity. Histogram and SteadyTime degenerate to near-constant alerting on the eight-feature subset and therefore produce identical aggregate scores. The interleaved supervised split is shown only as a diagnostic comparison because adjacent train and test rows share local scenario context.

\begin{table}[!t]
\caption{Baseline IDS results on the Data4Cyber state logs (eight-feature subset). Supervised detectors use the chronologically blocked split.}
\label{tab:ipal_eval}
\centering
\scriptsize
\setlength{\tabcolsep}{2.3pt}
\renewcommand{\arraystretch}{0.99}
\begin{tabular}{L{1.48cm} L{1.20cm} c c c c}
\toprule
\textbf{IDS} & \textbf{Train/Test} & \textbf{Prec.} & \textbf{Rec.} & \textbf{F1} & \textbf{FPR}\\
\midrule
\multicolumn{6}{l}{\emph{Unsupervised/process-aware (train S0 only)}}\\
MinMax & S0--all & 0.654 & 0.589 & 0.620 & 0.471\\
Histogram & S0--all & 0.703 & 1.000 & 0.825 & 0.640\\
SteadyTime & S0--all & 0.703 & 1.000 & 0.825 & 0.640\\
ExistsIDS & S0--all & 0.706 & 0.999 & 0.827 & 0.630\\
Dec.Places & S0--all & 0.753 & 0.208 & 0.327 & 0.103\\
IsolationF. & S0--all & 0.846 & 0.369 & 0.514 & 0.101\\
GeCo & S0--all & 0.876 & 0.522 & 0.654 & 0.111\\
\midrule
\multicolumn{6}{l}{\emph{Supervised (chronologically blocked split)}}\\
DecisionTree & blocked & 0.699 & 0.673 & 0.686 & 0.918\\
RandomForest & blocked & 0.746 & 0.796 & 0.771 & 0.858\\
ExtraTrees & blocked & 0.690 & 0.613 & 0.649 & 0.873\\
NaiveBayes & blocked & 0.839 & 0.789 & 0.813 & 0.481\\
SVM & blocked & 0.726 & 0.824 & 0.772 & 0.985\\
\bottomrule
\end{tabular}
\end{table}

\subsection{IDS Discussion}
\label{sec:discussion}
Three implications follow from Table~\ref{tab:ipal_eval} and Fig.~\ref{fig:ids_benchmark}. First, S0-trained baselines show the expected OT precision-recall trade-off: existence- and histogram-style detectors reach recall near one at false-positive rates of 0.63--0.64, whereas DecimalPlaces and Isolation~Forest keep false positives near 10\,percent but at lower recall. Second, interleaved evaluation gives overly optimistic supervised scores, but blocked F1 falls to 0.649--0.813 with elevated false-positive rates. Third, the process-aware baseline is the only S0-trained detector combining low false-positive rate with non-trivial recall, and its F1 is comparable to blocked Extra~Trees (0.654 vs.\ 0.649) at far lower false-positive rate (0.111 vs.\ 0.873).

The aggregate scores are consistent with the signal-level patterns in Fig.~\ref{fig:results_comparison}. S1 and S2 are comparatively subtle for S0-trained state detectors because the process response is localized to the manipulated PV or BSS path and can overlap with benign dispatch variation. S4 and S5 create clearer physical inconsistencies because FDI affects multiple process paths and creates stronger residuals between reported and reference state. S6 is different again: the manipulated variable is contextual, and the physical response must be interpreted through the EMS dispatch logic rather than through one process channel alone. The largest research headroom therefore lies with process-aware and physics-informed methods that cross-validate Modbus telemetry against the EMS dispatch model, exploit the reference-versus-reported SoC residual, and incorporate cable parameters and feeder topology.

\subsection{Experimental Scope}
\label{sec:limitations}
Data4Cyber is recorded in a controlled laboratory setting with vendor-specific Modbus and MQTT semantics. The contained topology and bounded 30-minute runs are deliberate: they keep cyber-physical cause-effect relations traceable and yield label-tight attack intervals. The scenario scope focuses on the OT-side manipulation and physical-integrity phase of the attack chain. IT-stage precursors such as phishing, credential theft, or ransomware delivery are outside the measured scenario scope; this choice isolates clean process signals for anomaly detection. The S6 scenario illustrates why this scope remains challenging: an adversary can cause physical impact through high-level price signals while leaving minimal overt traces in lower-level Modbus polling.

\subsection{Data Availability and Evaluation Recommendations}
For downstream studies, we recommend preserving chronological order unless the split is explicitly labeled as a parser or diagnostic check. S0 should be used for benign calibration and false-positive reporting, while attack scenarios should be evaluated with blocked or scenario-wise partitions. Results should report precision, recall, false-positive rate, and feature channels because the scenarios differ in observability: S1--S2 are process-subtle, S4--S5 are physically prominent, and S6 is context-driven. Data4Cyber is published on Zenodo at DOI \href{https://doi.org/10.5281/zenodo.19965384}{10.5281/zenodo.19965384}.

\ifdefined\FloatBarrier
\FloatBarrier
\fi

\section{Conclusion and Outlook}
\label{sec:conclusion}
Data4Cyber is a labeled cyber-physical dataset for DER-centric DMS environments with mixed Modbus/TCP and MQTT communication, combining synchronized process measurements, OT network captures, controller-relevant context, phase-resolved labels, and scenario metadata across one benign baseline and six attack scenarios. Baseline IDS evaluation shows that high recall alone is insufficient in this setting, because several detectors achieve sensitivity through alarm saturation, while context-driven attacks such as MQTT price manipulation require interpretation across control logic and physical state. Data4Cyber therefore provides an empirical basis for process-aware and physics-informed detectors that exploit the EMS dispatch model, the reference-versus-reported SoC residual, and feeder topology. Future extensions should broaden the benchmark with longer campaigns, diurnal and seasonal variation, additional attack chains, and larger asset sets.

\ifdefined\balance
\balance
\fi

\end{document}